\documentclass[preprint,showpacs,preprintnumbers,amsmath,amssymb]{article}

\usepackage{graphicx}
\usepackage{dcolumn}
\usepackage{bm}

\begin{document}

\title{Killing Reduction of 5-Dimensional Spacetimes}

\author{Xuejun Yang\footnote{yang\_xue\_jun@sohu.com},
Yongge Ma\footnote{yonggema@yahoo.com}, Jianbing Shao, and Wei Zhou\\
\small Department of Physics, Beijing Normal University, Beijing
100875, CHINA}

\date{\today}

\maketitle

\begin{abstract}
In a 5-dimensional spacetime ($M,g_{ab}$) with a Killing vector
field $\xi ^a$ which is either everywhere timelike or everywhere
spacelike, the collection of all trajectories of $\xi ^a$ gives a
4-dimensional space $S$. The reduction of ($M,g_{ab}$) is studied
in the geometric language, which is a generalization of Geroch's
method for the reduction of 4-dimensional spacetime. A
4-dimensional gravity coupled to a vector field and a scalar field
on $S$ is obtained by the reduction of vacuum Einstein's equations
on $M$ , which gives also an alternative description of the
5-dimensional Kaluza-Klein theory. Besides the symmetry-reduced
action from the Hilbert action on $M$, an alternative action of
the fields on $S$ is also obtained, the variations of which lead
to the same fields equations as those reduced from the vacuum
Einstein equation on $M$.

\end{abstract}

{PACS number(s): 04.50.+h, 04.20.Fy}

\section{Introduction}

   Spacetime reduction is very important in any high dimensional
theory of physics such as Kaluza-Klein
theory\cite{Duff}-\cite{Darabi}, high dimensional theory of
gravity\cite{Overduin} \cite{Weinberg} and string
theory\cite{Polchinski}-\cite{Yega}. The dimensional reduction can
make a high dimensional theory contact with the 4-dimensional
sensational world. It is also a useful approach to study
spacetimes with symmetries. The original Kaluza-Klein theory
unifies electromagnetic and gravitational interactions in four
dimensions by a 5-dimensional spacetime. Since the theory was
first proposed by Kaluza\cite{Kaluza}, it has been studied by a
large number of authors\cite{Duff}-\cite{Darabi} and also extended
to higher dimensions in order to give rise to a unification of
gravity with non-Abelian gauge theories\cite{Appelquist}. Along
with the successful description of the electroweak and strong
interactions as gauge theories, the K-K approach may serve as a
framework for studying their unification with gravity. Also using
the idea of supergravity in K-K theory, a geometric description of
both gauge fields and spinor matter is possible.

In ref.\cite{Geroch}, Geroch introduced a Killing reduction
formalism of 4-dimensional spacetime. Let ($\mathcal{M},g_{ab}$)
be a 4-dimensional spacetime with Killing vector field $\xi^a$
which is either everywhere timelike or everywhere spacelike. The
collection of all trajectories of $\xi^{a}$ gives a 3-dimensional
space $\Sigma$.
 The discussion of Geroch shows that there is a one-to-one
correspondence between tensor fields and tensor operations on
$\Sigma$ and the certain tensor fields and tensor operations on
$\mathcal{M}$. The differential geometry of $\Sigma$ will, in this
sense, be mirrored in $\mathcal{M}$. Geroch gives the relations of
geometric properties between $\Sigma$ and $\mathcal{M}$ and then
obtains the field equations on $\Sigma$, which describes a
3-dimensional gravity coupled to two massless scalar fields and is
equivalent to the vacuum Einstein equations on $\mathcal{M}$.

In this paper we extend Geroch's approach to a 5-dimensional
spacetime ($M,g_{ab}$) with a Killing vector field in order to
study its reduction. We first obtain a series of equations on $S$,
which is also the collection of all trajectories of $\xi^a$,
parallel to Geroch's. The results show that 5-dimensional gravity
with a Killing vector is equivalent to 4-dimensional gravity
$h_{ab}$ coupled to a vector field $A_a$ and a scalar field
$\lambda$, and hence it is consistent with the conclusion of the
Kaluza-Klein theory. We then study the reduction from the
viewpoint of variation principle. It turns out that the
symmetry-reduced action from the Hilbert action on $M$ would give
the correct reduced field equations only if one used $B_a$, which
has a trivial relation with the 5-metric components, rather than
$A_a$ as one of the arguments. Finally, we propose another
4-dimensional action on $S$. Its variations with respect to
$h^{ab}$, $\lambda$ and $A_a$ can give the same reduced field
equations.

\section{Symmetric Reduction of 5-Dimensional Spacetime}
 Let ($M,g_{ab}$) be a n-dimensional spacetime with a Killing vector field $\xi
^a$, which is either everywhere timelike or everywhere spacelike.
Let $S$ denote the collection of all trajectories of $\xi^a$. A
map $\psi$ from $M$ to $S$ can be defined as follows: For each
point p of $M$, $\psi(p)$ is the trajectory of $\xi^a$ passing
through p. Assume $S$ is given the structure of a differentiable
(n-1)-manifold such that $\psi$ is a smooth mapping. It is natural
to regard $S$ as a quotient space of $M$. The proof of Geroch
about the following conclusion is independent of the dimension of
$M$ \cite{Tavanfar}: There is a one-to-one correspondence between
tensor fields $\tilde{T}_{a...c}^{b...d}$ on $S$ and tensor fields
$T_{a...c}^{b...d}$ on $M$ which satisfy

\begin{equation}
\begin{array}{l}
\xi^a T^{b\cdots d}_{a\cdots c}=0,\ \cdots \ ,\xi_d T^{b\cdots
d}_{a\cdots
  c}=0 ,\\
{\mathcal{L}}_\xi T^{b\cdots d}_{a\cdots c}=0 . \label{Li D}
\end{array}
\end{equation}
The entire tensor field algebra on $S$ is completely and uniquely
mirrored by tensor field on $M$ subject to (\ref{Li D}). Thus, we
shall speak of tensor fields being on $S$ merely as a shorthand
way of saying that the fields on $M$ satisfy (\ref{Li D}).

The metric, inverse metric and the Kronecker delta on $S$ are
defined as
\begin{equation}
h_{ab}=g_{ab}-\lambda ^{-1}\xi _a\xi_b, \label{h_{ab}}
\end{equation}

\begin{equation}
h^{ab}=g^{ab}-\lambda ^{-1}\xi ^a\xi ^b,
 \label{h^{ab}}
\end{equation}

\begin{equation}
h_a^b=\delta _a^b-\lambda ^{-1}\xi _a\xi ^b,
 \label{h_a^b}
\end{equation}
where $\lambda \equiv\xi ^a\xi _a$. (\ref{h_a^b}) can also be
interpreted as the projection operator onto $S$. The covariant
derivative on $S$ is defined by
\begin{equation}
D_eT_{a\cdots c}^{b\cdots d}=h_e^ph_a^m\cdots h_c^nh_r^b\cdots
h_s^d\nabla _pT_{m\cdots n}^{r\cdots s},\label{D T}
\end{equation}
where $\nabla _p$ is the covariant derivative associated with the
metric $ g_{ab}$ on $M$ and $T_{a\cdots c}^{b\cdots d}$ is any
tensor field on $S$. Note that $D_e$ satisfies all the conditions
of a derivative operator and $D_c h_{ab}=0$.

We now consider the special case where n=5. It can be shown that
$\varepsilon _{abcd}\equiv |\lambda|^{-\frac{1}{2} }\varepsilon
_{abcde}\xi ^e$ is the volume element associated with the metric
$h_{ab}$ on $S$, i.e.,

\begin{equation}
D_f\varepsilon_{abcd}=0, \label{silon}
\end{equation}
here $\varepsilon _{abcde}$ is the volume element associated with
the metric $ g_{ab}$ on $M$, i.e.,
 $\nabla _f\varepsilon
_{abcde}=0.$ We define the twist 2-form $\omega_{ab}$ of $\xi ^a$
by
\begin{equation}
\omega_{ab}:=\varepsilon _{abcde}\xi^c\nabla ^d\xi^e.
\label{omega}
\end{equation}
Clearly we have $\omega_{ab}=\omega_{[ab]}$ and $\omega_{ab}\xi
^a=0$. It is also easy to see that
\begin{equation}
\mathcal{L} _\xi \lambda =0,\hspace{1cm} \mathcal{L} _\xi
\omega_{ab}=0. \label{Li lambda}
\end{equation}
Hence, $\lambda $ and $\omega_{ab}$ are fields on $S$.

Lengthy but straightforward calculations lead to the following
results.

1) The Riemann tensor $R_{abcd}$ of ($S,h_{ab}$) is related to the
Riemann tensor $ \mathcal{R}_{abcd}$ of ($M,g_{ab}$) by
\begin{equation}
R_{abcd}=h_{[a}^ph_{b]}^qh_{[c}^rh_{d]}^s[\mathcal{R}_{pqrs}+2\lambda
^{-1}(\nabla _p\xi _q)(\nabla _r\xi_s)+2\lambda ^{-1}(\nabla
_p\xi_r)(\nabla _q\xi_s)]; \label{R abcd}
\end{equation}

2)The derivative of the Killing vector reads
\begin{equation}
\nabla _a\xi _b=-\frac {1}{4}\lambda ^{-1}\varepsilon _{abcde}\xi
^c\omega^{de}+\lambda ^{-1}\xi _{[b}D_{a]}\lambda ;\label{nablaxi}
\end{equation}

3)The second derivative reads
\begin{equation}
\nabla _a\nabla _b\xi _c=\mathcal{R}_{dabc}\xi ^d;
\label{nablanablaxi}
\end{equation}

4) Contracting (\ref{R abcd}) and using (\ref{nablaxi}) and
(\ref{nablanablaxi}), the relation between the Ricci tensor
$R_{ab}$ of ($S,h_{ab}$) and the Ricci tensor $\mathcal{R}_{ab}$
of ($M,g_{ab}$) can be obtained as
\begin{equation}
R_{ab}=\frac{1}{2} \lambda ^{-2}(\omega_a^m\omega_{bm}-\frac{1}{2}
h_{ab}\omega_{mn}\omega^{mn})+\frac{1}{2}\lambda
^{-1}D_aD_b\lambda -\frac{1}{4}\lambda ^{-2}(D_a\lambda)
D_b\lambda +h_a^mh_b^n\mathcal{R}_{mn} ;\label{R_{ab}}
\end{equation}

5) Taking the curl and divergence of (\ref{omega}) and using
(\ref{nablaxi}) and (\ref{nablanablaxi}), one can get
\begin{equation}
D_{[a}\omega_{bc]}=\frac{2}{3}\varepsilon _{abcmn}\xi
^m\mathcal{R}_p^{n}\xi ^p, \label{Domega}
\end{equation}
and
\begin{equation}
D^a\omega_{ab}=\frac{3}{2}\lambda ^{-1}\omega_{mb}D^m\lambda;
\label{Daaomega}
\end{equation}

6) Using (\ref{nablaxi}) and (\ref{nablanablaxi}) again, we have
\begin{equation}
D^2\lambda =\frac 12\lambda ^{-1}(D^a\lambda) D_a\lambda -\frac
12\lambda ^{-1}\omega^{ab}\omega_{ab}-2\mathcal{R}_{mn}\xi ^m\xi
^n, \label{D^2lambda}
\end{equation}
where $D^2\equiv D^aD_a$;

7) Contracting (\ref{R_{ab}}) by $h^{ab}$ and using
(\ref{D^2lambda}), we obtain
\begin{equation}
R=-\frac 14\lambda ^{-2}\omega_{mn}\omega^{mn}+\lambda
^{-1}D^2\lambda -\frac 12\lambda ^{-2}(D^a\lambda) D_a\lambda
+\mathcal{R}, \label{R}
\end{equation}
here, $R$ and $\mathcal{R}$ are the scalar curvature of
($S,h_{ab}$) and ($M,g_{ab}$) respectively.

Thus, the basic equations for a 5-dimensional spacetime with a
Killing vector field are a set of differential equations on three
variables, the metric $h^{ab}$, the norm $\lambda $ and the twist
2-form $\omega_{ab}$ of the Killing vector. They
 are formulated as (\ref{R_{ab}}), (\ref{Domega}),
(\ref{Daaomega}) and (\ref{D^2lambda}).

 When the spacetime ($M,g_{ab}$) is source free ($\mathcal{R}_{ab}=0$),
(\ref{Domega}) implies $(d\omega)_{abc}$ on $S$ is zero. Hence at
least locally there is a 1-form $A_a$ on $S$ such that
$\omega_{ab}=(dA)_{ab} \equiv 2D_{[a}A_{b]}$. Then the equations
(\ref{R_{ab}})-(\ref{D^2lambda}) become

\begin{equation}
\begin{array}{lll}
R_{ab}&=&\frac 12\lambda ^{-2}[(dA)_a^m(dA)_{bm}-\frac
12h_{ab}(dA)_{mn}(dA)^{mn}]\\
 &&+\frac 12\lambda
^{-1}D_aD_b\lambda -\frac 14\lambda ^{-2}(D_a\lambda) D_b\lambda;
 \label{R_{ab}1}
\end{array}
\end{equation}

\begin{equation}
D^a(dA)_{ab}=\frac 32\lambda ^{-1}(dA)_{mb}D^m\lambda;
\label{D(ab)}
\end{equation}

\begin{equation}
D^2\lambda =\frac 12\lambda ^{-1}(D^a\lambda) D_a\lambda -\frac
12\lambda ^{-1}(dA)_{ab}(dA)^{ab}. \label{D^2lambda1}
\end{equation}
Equations (\ref{R_{ab}1})-(\ref{D^2lambda1}) describe a
4-dimensional gravity coupled to a vector field and a scalar
field, which are equivalent to 5-dimensional vacuum Einstein's
equation with a Killing symmetry. When the Killing vector field is
spacelike, it coincides with the conclusion of the 5-dimensional
Kaluza-Klein theory. Thus, by reducing Einstein's equation on $M$
by an extension of  Geroch's method, we can provide an alternative
description of the Kaluza-Klein theory in 4+1 dimensions.

\section{Symmetry-Reduced Hilbert Action on $S$ }

For practical calculations, it is convenient to take a coordinate
system adapted to the congruence:

\begin{equation}
(\frac \partial {\partial x^5})^a=\xi ^a .\label{x5}
\end{equation}
From (\ref{h_{ab}}), we have

\begin{equation}
g_{ab}=h_{ab}+\lambda ^{-1}\xi _a\xi _b. \label{g_{ab}h_{ab}}
\end{equation}
The components of (\ref{g_{ab}h_{ab}}) are

\begin{equation}
g_{\mu\nu} =h_{\mu\nu}+\lambda ^{-1}\xi _{\mu}\xi
_{\nu},\hspace{1cm} \mu, \nu =1,...,5, \label{g_{munu}}
\end{equation}
Particularly one has
\begin{equation}
g_{\mu 5}=g_{ab}(\frac \partial {\partial x^{\mu}})^a\xi ^b=(\frac
\partial
{\partial x^{\mu}})^a\xi _a=\xi _{\mu}\label{g_{mu5}}
\end{equation}
and
\begin{equation}
g_{55}=g_{ab}\xi ^a\xi ^b=(\frac \partial {\partial x^5})^a\xi
_a=\xi _5=\lambda. \label{g_{55}}
\end{equation}
Let
\begin{equation}
B_a^{\prime }=\lambda ^{-1}\xi _a-(dx^5)_a, \label{Bprime}
\end{equation}
then $\xi ^aB_a^{\prime }=0$. One can also prove that $\mathcal{L}
_\xi B_a^{\prime }=0$, hence $B_a^{\prime }$ is a 1-form on $S$.
Note that $B_a^{\prime }$ is dependent of the coordinate system
chosen. Using (\ref{nablaxi}) and (\ref{Bprime}), it can be shown
that
\begin{equation}
(dB^{\prime })_{ba}\equiv 2D_{[b}B_{a]}^{\prime }=\frac 12|
\lambda |^{-\frac 32}\varepsilon _{abcd}\omega^{cd}\equiv
\bar{\omega}_{ba}. \label{dBprime}
\end{equation}
Eq. (\ref{dBprime}) means $(d\bar{\omega})_{cba}=0$, hence there
is at least locally a one-form $B_a$ on $S$ such that
$\bar{\omega}_{ba}=(dB)_{ba}$. This definition of $B_a$ is
independent of any coordinate system and hence purely geometric.
Thus, we obtain
\begin{equation}
\begin{array}{lll}
(dB)_{ba}&=&\frac 12|\lambda|^{-\frac 32}\varepsilon
_{abcd}\omega^{cd},\\
\hspace{0.5cm}\omega_{ab}&=&\pm \frac 12|\lambda|^{\frac
32}\varepsilon _{abcd}(dB)^{cd}.
\end{array}
\label{dB}
\end{equation}
Hereafter, when $\lambda >0$ or $\lambda <0$, the sign ''$\pm $''
means ''+'' or ''-'' respectively. The Hilbert action on $M$ reads
\begin{equation}
S[g^{ab}]=\int_M\sqrt{-g}\mathcal{R} .\label{Sg}
\end{equation}
Since the principle of symmetric criticality is valid in our one
Killing vector model \cite{Fels}, one expects that the reduced
field Eqs. (\ref{R_{ab}1})-(\ref{D^2lambda1}) could be obtained by
the variation of the action from the symmetric reduction of
(\ref{Sg}). Using (\ref{g_{munu}}), we obtain
\begin{equation}
g=\lambda h, \label{gh}
\end{equation}
where $g$ and $h$ are respectively the determinants of components
$g_{\mu\nu} (\mu, \nu=1,2,...5)$ and $h_{\mu\nu} (\mu,
\nu=1...4)$. Using (\ref{R}) and (\ref{gh}), the action (\ref{Sg})
on $M$ is reduced to the following action on $S$ up to a boundary
term

\begin{equation}
S[h^{ab},\lambda ,\omega_{ab}]=\int_S|\lambda |^{\frac
12}\sqrt{|h|}[ R +\frac 14\lambda ^{-2}w_{mn}w^{mn}].
\label{Shomega}
\end{equation}
The variation of action (\ref{Shomega}) with respect to ($h^{ab}$,
$\lambda$, $\omega_{mn}$) can not give the correct reduced field
equations. In order to get those equations, one has to use the
5-metric components rather than $\omega_{mn}$ as the arguments of
the reduced action. This subtlety exists also in the symmetric
reduction of 4-dimensional spacetimes\cite{HeMaYang}.

From (\ref{dB}), we have
\begin{equation}
\omega^{mn}\omega_{mn}=-\lambda ^3(dB)_{ab}(dB)^{ab}.
\label{omega^{mn}omega_{mn}}
\end{equation}
Substituting (\ref{omega^{mn}omega_{mn}}) into (\ref{Shomega}), we
obtain the action on $S$ in terms of basic variables $h^{ab}$,
$\lambda $ and $B_a$ as
\begin{equation}
S[h^{ab},\lambda ,B_a]=\int_S|\lambda|^{\frac 12}\sqrt{|
h|}[R-\frac 14\lambda (dB)_{ab}(dB)^{ab}], \label{ShB}
\end{equation}
which has the same form as the reduced action in 5-dimensional K-K
theory. The variations of this action with respect to ($h^{ab}$,
$\lambda$, $B_c$) will give the correct reduced fields equations
on $S$.
 Especially, the variation of action (\ref{ShB})
with respect to $B_a$ gives
\begin{equation}
D^c(dB)_{ac}=-\frac 32\lambda ^{-1}(D^c\lambda) (dB)_{ac}.
\label{DdB}
\end{equation}
Substituting (\ref{dB}) into (\ref{DdB}), we get
\begin{equation}
D_{[a}\omega_{bc]}=0 \label{Dabcomega}.
\end{equation}
Eq. (\ref{Dabcomega}) means that there is at least locally a
1-form $A_a$ on $S$ such that
\begin{equation}
\omega_{ab}=(dA)_{ab}. \label{omega1}
\end{equation}
Substituting $A_a$ for $B_a$ in Eq. (\ref{DdB}) through Eqs.
(\ref{omega1}) and (\ref{dB}), we obtain Eq. (\ref{D(ab)}). Eqs.
(\ref{R_{ab}1}) and (\ref{D^2lambda1}) can also be obtained by
substituting $A_a$ for $B_a$ after the corresponding variations.

\section{An Alternative Action on $S$ }

If one wants to take $h^{ab}$, $\lambda $ and $A_a$ as basic
variables, action (\ref{Shomega}) fails to be the right action.
Another action is thus needed. When the Killing field $\xi ^a$ is
hypersurface orthogonal, from Eqs. (\ref{g_{mu5}}),
(\ref{g_{55}}), (\ref{Bprime}), (\ref{dBprime}) and (\ref{R}), we
have

\begin{equation}
\omega_{ab}=0 , \label{omega2}
\end{equation}

\begin{equation}
\mathcal{R}=R-\lambda ^{-1}h^{ab}D_aD_b\lambda +\frac 12\lambda
^{-2}h^{ab}(D_a\lambda) D_b\lambda .\label{tildeR5}
\end{equation}
In this case, up to a boundary term the Hilbert action (\ref{Sg})
on $M$ is reduced on $S$ as \cite{Fels}

\begin{equation}
S[h^{ab},\lambda ]=\int_S|\lambda|^{\frac 12}\sqrt{|h|}R.
\label{Shlambda}
\end{equation}
To obtain a regular form of action on $S$, we conformally
transform $h^{ab}$ as
\begin{equation}
\tilde{h}_{ab}=\Omega ^{-2}h_{ab}. \label{tildeh}
\end{equation}
Let $\tilde{D}_a$ be the covariant derivative operator determined
by metric $\tilde{h}_{ab}$, i.e.,
\begin{equation}
\tilde{D}_a\tilde{h}_{bc}=0. \label{tildeDtildeh}
\end{equation}
The relation of $\tilde{D}_a$ and $D_a$ is \cite{Wald}
\begin{equation}
\begin{array}{lll}
D_a v_b =\tilde{D}_a v_b-C_{ab}^c
v_c,\hspace{1cm}\forall\hspace{0.3cm} v_a \in \mathcal{F} _S(0,1),
\end{array}
\label{Dv}
\end{equation}
where
\begin{equation}
\begin{array}{lll}
C_{ab}^c &=&\frac
12h^{cd}(\tilde{D}_ah_{bd}+\tilde{D}_bh_{ad}-\tilde{D}_dh_{ab})
\\
&=&\tilde{h}_b^c\tilde{D}_a\ln \Omega +\tilde{h}_a^c\tilde{D}_b\ln
\Omega -\tilde{h} _{ab}\tilde{h}^{cd}\tilde{D}_d\ln \Omega.
\end{array}
\label{C_{ab}^c}
\end{equation}
Let
\begin{equation}
\Omega =|\lambda|^{-\frac14}. \label{Omega1}
\end{equation}
Ignoring the boundary term (\ref{Shlambda}) becomes
\begin{equation}
S[\tilde{h}^{ab},\lambda
]=\int_S\sqrt{|\tilde{h}|}[\tilde{R}-\frac 38\lambda ^{-2}
\tilde{h}^{ab}(\tilde{D}_a\lambda) \tilde{D}_b\lambda ].
\label{Stildehlambda1}
\end{equation}
Let $\Lambda =\sqrt{6}\ln \Omega $, then (\ref{Stildehlambda1})
becomes
\begin{equation}
S[\tilde{h}^{ab},\Lambda
]=\int_S\sqrt{|\tilde{h}|}[\tilde{R}-\tilde{h}^{ab}(\tilde{D}
_a\Lambda) \tilde{D}_b\Lambda ].\label{StildehLambda}
\end{equation}
This is the action of a 4-gravity $\tilde{h}^{ab}$ coupled to a
massless scalar field. Thus in the source-free case, a
5-dimensional spacetime with a hypersurface orthogonal Killing
vector field which is either everywhere timelike or everywhere
spacelike is "comformally" equivalent to 4-dimensional gravity
coupled to a massless Klein-Gordon field.

In general case, from Eqs. (\ref{R_{ab}1})-(\ref{D^2lambda1}) we
know that besides a scalar field $ \lambda $, the 4-gravity
couples also to a vector field $A_a$ on $S$. By carefully
observing the reduced field equations and the special action
(\ref{Stildehlambda1}), we suppose the following action being the
one we are looking for,
\begin{equation}
S[\tilde{h}^{ab},\lambda
,A_a]=\int_S\sqrt{|\tilde{h}|}[\tilde{R}-\frac 38\lambda
^{-2}\tilde{h}^{ab}(\tilde{D}_a\lambda) \tilde{D}_b\lambda -\frac
14|\lambda|^{-
\frac32}\tilde{h}^{ac}\tilde{h}^{bd}(dA)_{ab}(dA)_{cd}].
\label{ShA}
\end{equation}
Varying the action (\ref{ShA}) with respect to $\tilde{h}^{ab}$,
$\lambda $ and $A_a$ respectively, we get
\begin{equation}
\begin{array}{lll}
\tilde{R}_{ab}&=&\frac 38\lambda ^{-2}(\tilde{D}_a\lambda)
\tilde{D}_b\lambda -\frac 18|\lambda
|^{-\frac32}\tilde{h}_{ab}\tilde{h}^{ec}\tilde{h}
^{fd}(dA)_{ef}(dA)_{cd}\\
&&+\frac 12|\lambda|^{-\frac32}\tilde{h} ^{cd}(dA)_{ac}(dA)_{bd};
\end{array}
\label{tildeRab1}
\end{equation}

\begin{equation}
\tilde{h}^{ab}\tilde{D}_a\tilde{D}_b\lambda =\lambda
^{-1}\tilde{h}^{ab}(\tilde{D} _a\lambda) \tilde{D}_b\lambda +\frac
12|\lambda|^{-\frac12}\tilde{h}^{ac}\tilde{
h}^{bd}(dA)_{ab}(dA)_{cd}; \label{tildehDaDb}
\end{equation}

\begin{equation}
D^a(dA)_{ab}=\frac 32\lambda ^{-1}(dA)_{ab}D^a\lambda. \label{DdA}
\end{equation}
Making the conformal transformation (\ref{tildeh}) inversely, Eqs.
(\ref{tildeRab1}) and (\ref{tildehDaDb}) become

\begin{equation}
\begin{array}{lll}
R_{ab}&=&\frac 12\lambda ^{-2}[h^{cd}(dA)_{ac}(dA)_{bd}-\frac
12h_{ab}h^{ec}h^{fd}(dA)_{ef}(dA)_{cd}]\\
&&+\frac 12\lambda ^{-1}D_aD_b\lambda -\frac 14\lambda
^{-2}(D_a\lambda) D_b\lambda;
\end{array}
\label{Rab}
\end{equation}

\begin{equation}
D^2\lambda =\frac 12\lambda ^{-1}h^{ab}(D_a\lambda) D_b\lambda
-\frac 12\lambda ^{-1}h^{ac}h^{bd}(dA)_{ab}(dA)_{cd}.
\label{D^2lambda3}
\end{equation}
Eqs. (\ref{Rab}), (\ref{DdA}) and (\ref{D^2lambda3}) are exactly
the same as (\ref{R_{ab}1})-(\ref{D^2lambda1}). Therefore, the
action (\ref{ShA}) is just the one which gives also the right
reduced field equations.

In conclusion, we have studied the symmetric reduction of
5-dimensional spacetime in three hands. First, we obtain
4-dimensional gravity coupled to a vector field and a scalar field
on $S$ by a direct reduction of vacuum Einstein's equation on $M$.
Then, taking $h^{ab}$, $\lambda $ and $B_a$ as basic variables, we
get reduced field equations by varying the symmetry-reduced action
on $S$, which is obtained by the reduction of the Hilbert action
on $M$. Finally, we propose an alternative action on $S$, which
allows us to take $h^{ab}$, $\lambda $ and $A_a$ as basic
variables, and its variations gives also the right reduced field
equations. All discussions are geometrically presented. The scheme
might also be extended to higher-dimensional Kaluza-Klein theories
that attempt to unify gauge theories with gravity.

\section*{ Acknowledgments}

This work is  supported in part by NSFC (10205002), YSRF for ROCS,
SEM, and Young Teachers Foundation of BNU. X. Yang would also like
to acknowledge support from NSFC (10073002).


\begin{thebibliography}{99}

\bibitem{Duff} M. J. Duff and B. E. W. Nilsson, Phys. Rep. {\bf 130}, 1 (1986).

\bibitem{Dimopoulos} N. Arkani-Hamed, S. Dimopoulos and
          G. Dvali, Phys. Lett. B{\bf 429}, 263 (1998); Phys. Rev. D{\bf 59}, 086004 (1999).

\bibitem{Han} T. Han, J. D. Lykken and R. Zhang, Phys. Rev. D{\bf 59}, 105006 (1999).

\bibitem{Giudice}G. F. Giudice, R. Rattazzi and J. D. Wells, Nucl. Phys. B{\bf 544}, 3(1999).

\bibitem{Darabi}F. Darabi and P. S. Wesson, Phys. Lett. B{\bf 527}, 1(2002).

\bibitem{Overduin}J. M. Overduin and P. S. Wesson, Phys. Rep. {\bf 283}, 305(1997).

\bibitem{Weinberg}S. Weinberg, Phys. Lett. B{\bf 125}, 265(1983).

\bibitem{Polchinski}J. Polchinski, String Theory I, II (Cambridge University
Press, Cambridge, 1998).

\bibitem{Antoniadis} I. Antoniadis, N. Arkani-Hamed, S. Dimopoulos and
          G. Dvali, Phys. Lett. B{\bf 436}, 257 (1998).

\bibitem{Yega}H. J. de Vega and N. Sanchez, Phys. Lett. B{\bf 197}, 320(1987).

\bibitem{Kaluza}T. Kaluza, Sitzungsber, Preuss. Akad. Wiss. Phys. Mat. Klasse, 966(1921).

\bibitem{Appelquist}T. Appelquist, A. Chodos and P. G. O. Freund, Modern Kaluza-Klein
Theories, Addison-Wesley Publishing Company, Menlo Park
,California(1987).

\bibitem{Geroch}R. Geroch, Jour. Math. Phys. {\bf 12}, 918(1971).

\bibitem{Tavanfar}M. Nouri-Zonoz and A. R. Tavanfar, JHEP {\bf 0302}, 059
(2003).

\bibitem{Fels}M. E. Fels and C. G. Torre, Class. Quant. Grav. {\bf 19}, 641(2002).

\bibitem{HeMaYang} H. He, Y. Ma, and X. Yang, Actions for vacuum Einstein's
equation with a Killing symmetry, gr-qc/0212101, accepted for
publication in Int. J. Mod. Phys. D.

\bibitem{Wald}  R. M. Wald, General relativity, (The University of Chicago Press,
1984).

\end{thebibliography}
\end{document}